\documentclass[twocolumn,aps,showpacs]{revtex4}

\usepackage{bm}
\usepackage{color}
\usepackage{graphicx}
\usepackage{dcolumn}
\usepackage{simplewick}

\begin{document}

\title{Relativistic coupled-cluster calculations of $^{20}$Ne, $^{40}$Ar, 
       $^{84}$Kr and $^{129}$Xe: correlation energies and dipole 
       polarizabilities}
\author{B. K. Mani and D. Angom}
\affiliation{Physical Research Laboratory,
             Navarangpura-380009, Gujarat, 
             India}

\begin{abstract}
   We have carried out a detailed and systematic study of the correlation 
   energies of inert gas atoms Ne, Ar, Kr and Xe using relativistic 
   many-body perturbation theory and relativistic coupled-cluster theory.
   In the relativistic coupled-cluster calculations, we implement
   perturbative triples and include these in the correlation energy 
   calculations. We then calculate the dipole polarizability of the 
   ground states using perturbed coupled-cluster theory.
\end{abstract}



\maketitle


\section{Introduction}

  High precision atomic experiments are at the core of several investigations 
into fundamental physics and high end technology developments. Selected
examples are search for electric dipole moment (EDM) \cite{griffith-09} and 
observation of parity nonconservation \cite{tsigutkin-09}. These endeavours, 
in general, require precision atomic theory calculations to analyse the 
results and understand systematics. The challenging part of precision atomic 
structure and properties calculations is obtaining accurate wave functions. In 
the case of high $Z$ atoms, the need to incorporate relativity adds to the 
difficulty. A systematic study of the correlation energy is one of the 
possible methods to test the accuracy of the atomic wave function. In this 
paper, we report the results of correlation energy calculations of inert gas 
atoms Ne, Ar, Kr and Xe. For this we employ many body perturbation theory 
(MBPT) and calculate the second order correlation energy. A comparative study 
reveals the changing nature of electron correlations in the group. Our 
interest in particular is Xe, which is a candidate for EDM 
experiments \cite{rosenberry-01} and theoretical calculations \cite{dzuba-02}.

   For completeness, in the presentation of the paper, we give an overview of 
MBPT. It is a powerful theory and forms the basis of other more sophisticated 
and elaborate many-body methods. However, one drawback of MBPT is the 
complexity of expressions at higher orders. This renders the theory
inappropriate to incorporate strong correlation effects in heavy atoms. Yet,
at lower orders its simplicity makes it an ideal choice to test and optimize
basis sets. We use this insight to generate basis sets for coupled-cluster
calculations.
 
  The coupled-cluster theory, first developed in nuclear many body physics
\cite{coester-58,coester-60}, is considered the most accurate many body 
theory. In recent times, it has been used with great success
in nuclear \cite{hagen-08}, atomic \cite{nataraj-08,pal-07}, molecular 
\cite{isaev-04} and condensed matter \cite{bishop-09} calculations. 
It is equivalent to incorporating electron correlation effects to all 
orders in perturbation. The theory has been used in performing
high precision calculations to study the atomic structure and properties.
These include atomic electric dipole moments \cite{nataraj-08,latha-09}, 
parity nonconservation \cite{wansbeek-08},  hyperfine structure constants
\cite{pal-07,sahoo-09} and electromagnetic transition properties 
\cite{thierfelder-09,sahoo-09a}. In the present work we use the relativistic
coupled-cluster singles and doubles (CCSD) approximation to calculate 
correlation energy and dipole polarizability of inert gas atoms Ne, Ar, Kr 
and Xe. In the dipole polarizability calculations, the dipole interaction 
Hamiltonian is introduced as a perturbation. A modified theory, recently 
developed \cite{latha-08}, incorporates the perturbation within the 
coupled-cluster theory. This theory has the advantage of subsuming correlation 
effects more accurately. The results provide a stringent test on the quality 
of the wave functions as the dipole polarizability of inert gas atoms are 
known to high accuracy \cite{hohm-90}. Based on the CCSD method, we also 
estimate the third order correlation energy. Further more, perturbative 
triples are incorporated in the coupled-cluster calculations.

  In the paper we give a brief description of MBPT in Section.II and
discuss the method to calculate electron correlation energy to the second 
and third order in residual Coulomb interaction. The coupled-cluster theory is 
described in Section.III, where we also discuss linearized coupled-cluster 
theory and correlation energy calculation using coupled-cluster theory. Then 
the inclusion of approximate triples to the correlation energy is explained 
and illustrated. Section.IV is a condensed description of the perturbed 
coupled-cluster theory and provide details of how to incorporate the effects 
of an additional perturbation to the residual Coulomb interaction in 
atomic systems. Results are presented and discussed in Section.V. In the 
paper, all the calculations and mathematical expressions are in atomic 
units ($e=\hbar=m_e=1$).


\section{Correlation Energy from MBPT}

  In this section, to illustrate the stages of our calculations and 
compare with coupled-cluster theory, we provide a brief description of 
many-body perturbation theory. Detailed and complete exposition of the method,
in the context of atomic many-body theory, can be found in ref
\cite{lindgren-85}.

 The Dirac-Coulomb Hamiltonian $H^{\rm DC}$ is an appropriate choice to 
incorporate relativistic effects in atoms. This is particularly true for heavy 
atoms, where the relativistic effects are large for the inner core electrons 
due to the high nuclear charge. As the name indicates, $H^{\rm DC}$ is fully 
relativistic for the one-body terms only. For an $N$ electron atom
\begin{equation}
  H^{\rm DC}=\sum_{i=1}^N\left [c\bm{\alpha}_i\cdot \bm{p}_i+
             (\beta-1)c^2 - V_N(r_i)\right ] +\sum_{i<j}\frac{1}{r_{ij}},
  \label{dchamil}
\end{equation}
where $\alpha_i$ and $\beta$ are the Dirac matrices. For the nuclear potential
$V_N(r)$, we consider the nucleus as finite size and the volume 
effects are taken into account by modeling the nuclear charge distribution 
as a Fermi two-parameter distribution. Then the nuclear density is 
\begin{equation}
  \rho_{\rm nuc}(r) = \frac{\rho_0}{1 + e^{\frac{(r-c)}{a}} },
\end{equation}
here, $a = t 4\ln 3$. The parameter $c$ is the half-charge radius, that is
$\rho_{\rm nuc}(c)=\rho_0/2$ and $t$ is the skin thickness. The eigen states of 
$H^{\rm DC}$ are $|\Psi_i\rangle$, the correlated many-particle states
with eigenvalues $E_i$. The eigenvalue equation is 
\begin{equation}
  H^{\rm DC}|\Psi_i\rangle = E_i|\Psi_i\rangle.
  \label{dcpsi}
\end{equation}
It is however impossible to solve this equation exactly due to the relative 
coordinates in the electron-electron Coulomb interaction. Many-body 
perturbation theory is one approach which, starting from a mean field
approximation, incorporates the electron correlation effects systematically.

The starting point of perturbative scheme in many-body theory is to split 
the Hamiltonian as
\begin{equation}
   H^{\rm DC} = H_0 + V,
\end{equation}
where $H_0 = \sum_i[c\alpha_i.p_i+(\beta_i-1)c^2-V_N{r_i}+u(\bm{r}_i)]$,
is the unperturbed or zeroth order Hamiltonian. It is the exactly solvable
part of the total Hamiltonian and correspond to independent particle model. 
In this model, each electron is assumed to move independently of the others 
in an average field arising from the nucleus and other electrons. The average 
field of the other electrons is the Dirac-Fock central potential $u(\bm{r}_i)$.
The remaining part of the electron-electron Coulomb interaction 
$V=\sum_{i<j}^N\frac{1}{\bm{r}_{ij}}-\sum_i u(\bm{r}_i)$, is the residual
Coulomb interaction. The purpose of any atomic many-body theory is to 
account for this part as accurately as possible. The Hamiltonian $H_0$ 
satisfies the eigenvalue equation
\begin{equation}
  H_0\vert\Phi_i\rangle=E_i^0\vert\Phi_i\rangle,
  \label{h0phi}
\end{equation}
where $\vert\Phi_i\rangle$ is a many-particle state and $E_i^0$ is the 
eigenvalue. The eigenstates are generally Slater determinants, antysymmetrised 
direct product of single particle states and $E_i^0$ is the sum of 
the single particle energies. The difference between the exact and mean field
energy, $\Delta E_i = E_i - E_i^0$, is the correlation energy of the 
$i^{\rm Th}$ state. At the single particle level, the relativistic spin
orbitals are of the form
\begin{equation}
  \psi_{n\kappa m}(\bm{r})=\frac{1}{r}
  \left(\begin{array}{r}
            P_{n\kappa}(r)\chi_{\kappa m}(\bm{r}/r)\\
           IQ_{n\kappa}(r)\chi_{-\kappa m}(\bm{r}/r)
       \end{array}\right),
\end{equation}
where $P_{n\kappa}(r)$ and $Q_{n\kappa}(r)$ are the large and small component
radial wave functions, $\kappa$ is the relativistic total angular momentum
quantum number and $\chi_{\kappa m}(\bm{r}/r)$ are the spin or spherical
harmonics. One representation of the radial components is to define these
as linear combination of Gaussian like functions and are referred to as
Gaussian type orbitals (GTOs). Then, the large and small 
components \cite{mohanty-89,chaudhuri-99} are
\begin{eqnarray}
   P_{n\kappa}(r) = \sum_p C^L_{\kappa p} g^L_{\kappa p}(r),  \nonumber \\
   Q_{n\kappa}(r) = \sum_p C^S_{\kappa p} g^S_{\kappa p}(r).
\end{eqnarray}
The index $p$ varies over the number of the basis functions. 
For large component we choose 
\begin{equation}
  g^L_{\kappa p}(r) = C^L_{\kappa i} r^{n_\kappa} e^{-\alpha_p r^2},
\end{equation}
here $n_\kappa$ is an integer. Similarly, the small component are 
derived from the large components using kinetic balance condition. The 
exponents in the above expression follow the general relation
\begin{equation}
  \alpha_p = \alpha_0 \beta^{p-1}.
  \label{param_gto}
\end{equation}
The parameters $\alpha_0$ and $\beta$ are optimized for an atom to provide 
good description of the atomic properties. In our case the optimization is to 
reproduce the numerical result of the total and orbital energies. 
Besides GTO, B-splines is another class of basis functions widely used in
relativistic atomic many body calculations \cite{johnson-88}. A description 
of B-splines with details of implementation and examples are given in 
ref \cite{johnson-07}.

 The next step in perturbative calculations is to divide the 
entire Hilbert space of $H_0$ into two manifolds: model and complementary 
spaces $P$ and $Q$ respectively. The model space has, in single reference 
calculation, the eigen state $|\Phi_i\rangle$ of $H_0$ which is a good 
approximation of the exact eigenstate $|\Psi_i\rangle$ to be calculated. The 
other eigenstates constitute the complementary space. The corresponding 
projection operators are defined as
\begin{equation}
  P = |\Phi_i\rangle\langle\Phi_i|
       \;\;\;\; {\rm and} \;\;\;\;
  Q = \sum_{|\Phi_j\rangle\notin P} |\Phi_j\rangle\langle\Phi_j|.
\end{equation}
The operator $P$ projects out the component of the exact eigenstate which
lies in the model space, $P\vert\Psi_i\rangle = |\Phi_i\rangle$ and $Q$
projects out the component in the orthogonal space and  $ P + Q = 1 $.
In the present paper, we restrict to calculating the ground state
$|\Psi_0\rangle$ of the closed shell inert gas atoms. From here on, for
a consistent description, the model space consist of $|\Phi_0\rangle$. 

 The most crucial part of perturbation theory is to define a wave operator 
$\Omega$ which operates on $|\Phi_0\rangle$  and transform it to 
$|\Psi_0\rangle$ as
\begin{equation}
  |\Psi_0\rangle=\Omega\vert\Phi_0\rangle .
\end{equation}
Then, with the intermediate normalization approximation
$\langle\Psi_0|\Phi_0\rangle = 1 $, the wave operator is evaluated in orders 
of the perturbation as $\Omega = \sum_{i=0}^\infty\Omega^{(i)}$ with 
$\Omega^{(0)} = 1 $.
It is possible to evaluate $\Omega^{(i)}$ iteratively or recursively from the
Bloch equation
\begin{equation}
 [\Omega,H_0]P = QV\Omega P-\chi PV\Omega P,
 \label{bloch} 
\end{equation}
where $\chi=\sum_{i=1}^\infty\Omega^{(i)}$ is the correlation operator. In the 
second quantization notations, the wave operator and perturbation can be 
expressed in terms of particle excitations. Then, the effect of correlation 
is incorporated as linear combination of excited states. For simplification, in 
the normal form the perturbation is separated as \cite{lindgren-85} 
$ V = V_0 + V_1 +V_2$. These zero-, one- and two-body operators and are 
defined as
\begin{eqnarray}
  V_0 && = \sum^{core}_a\langle a|-u|a\rangle + \frac{1}{2}\sum^{core}_{ab}
        (\langle ab|r^{-1}_{ab}|ab\rangle - \langle ba|r^{-1}_{ab}|ab\rangle),
         \nonumber \\
  V_1 &&= \sum_{ij}\{a_i^\dagger a_j\}\langle i|v|j\rangle,  \nonumber \\
  V_2 &&= \frac{1}{2}\sum_{ijkl}\{a_i^\dagger a_j^\dagger a_l a_k \}
          \langle ij|r_{12}^{-1}|kl\rangle.
\end{eqnarray}
The operators $a^{\dagger}$ ($a$) create (annihilate) electrons in virtual 
($p,q, r, s, ....$ etc) and core ($a, b, c, d, ....$ etc) shells. The indexes
$i, j, k, l$ etc are general representations of orbitals, it could either be
core or virtual. The operator $V_0$ acting on $|\Phi_0\rangle$ leaves it 
unchanged, while $V_1$ and $V_2$ produce single and double excitations. From 
these definition, the first-order wave operator can be separated as
\begin{equation}
  \Omega^{(1)} = \Omega^{(1)}_1 +\Omega^{(1)}_2 .
\end{equation}
Here, $\Omega^{(1)}_1$ and $\Omega^{(1)}_2$ are one- and two-body components
of the first order wave operator. The corresponding algebraic expressions 
are
\begin{eqnarray}
  \Omega^{(1)}_1 & = & \sum_{ap}a^\dagger_p a_a 
                       \frac{\langle p|v|a\rangle}{\epsilon_a-\epsilon_p}, 
                       \\
  \Omega^{(1)}_2 & = & \frac{1}{2}\sum_{abpq}a^\dagger_pa^\dagger_q a_ba_a
                       \frac{\langle pq|v|ab\rangle}{\epsilon_a + 
                       \epsilon_b - \epsilon_p - \epsilon_q}.
\end{eqnarray}
We get singly (doubly) excited states $|\Phi_a^p\rangle$ 
($|\Phi{ab}^{pq}\rangle$) when $\Omega^{(1)}_1$ ($\Omega^{(1)}_1$ ) operates
on the reference state $|\Phi_0$. The complexity of the expressions increases 
with order of perturbation and is hard to manage. One powerful tool in 
many-body perturbation theory is the diagrammatic evaluation of the 
perturbation expansion. Then, the tedious algebraic evaluations are reduced 
to a sequence of diagrams and equivalent algebraic expressions are derived 
with simple rules. Even with this approach, it is computationally not 
practical to go beyond fourth order.
%
%
\begin{center}
\begin{figure}[h]
  \includegraphics[width=5cm]{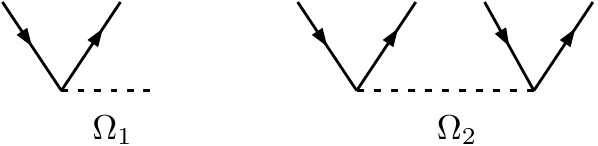}
  \caption{ The diagrammatic representations of the one- and two-body wave
            operator. Lines with downward (upward) arrows represent 
            core (virtual) single particle states. }
\end{figure}
\end{center}
%
%


\subsection{Second-Order Correlation Energy}

\begin{center}
\begin{figure}[h]
  \includegraphics[width = 5.0 cm]{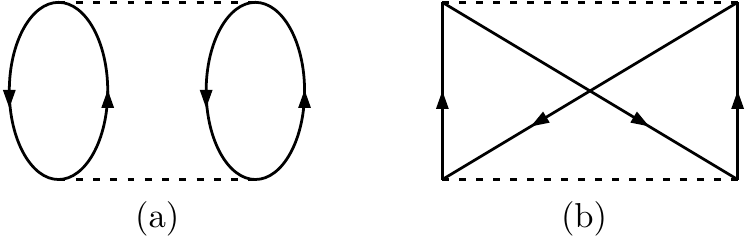}
  \caption{MBPT diagrams arising from the second-order correlation energy.}
  \label{2nd_e}
\end{figure}
\end{center}

 The ground state correlation energy $\Delta E_0$, as described earlier, is 
the difference between the exact energy and the mean field energy. It is the 
sum total of the energy corrections from all orders in perturbation. At the 
$n^{\rm th}$ order, the energy correction 
$E_{\rm corr}^{(n)}= \langle \Phi_0|V\Omega^{(n-1)}|\Phi_0\rangle$ and
$\Delta E_0 = \sum_n E_{\rm corr}^{(n)} $. Then the second order correlation
energy is
\begin{equation}
  E^{(2)}_{\rm corr} = \langle\Phi_0|(V_1 + V_2)(\Omega^{(1)}_1
            + \Omega^{(1)}_2)|\Phi_0\rangle.
\end{equation}
When Dirac-Fock orbitals are used, the diagonal matrix elements of $V_1$ 
are the orbital energies and off diagonal matrix elements are zero. For this
reason, it does not contribute to the second-order energy. Then, the second
order correlation energy is
\begin{equation}
  E^{(2)}_{\rm corr} = \langle\Phi_0|V_2\Omega^{(1)}_2|\Phi_0\rangle.
\end{equation}
There are two diagrams arising from the above expression and these are as 
shown in Fig.\ref{2nd_e}. 
%
%
In terms of algebraic expressions
\begin{equation}
  E^{(2)}_{\rm corr} = \sum_{abpq}\left [ \frac{\langle ab|V_2|pq\rangle
                       \langle pq|V_2|ab\rangle}{\epsilon_a + \epsilon_b -
                       \epsilon_p - \epsilon_q} - \frac{\langle ba|V_2|pq\rangle
                       \langle pq|V_2|ab\rangle}{\epsilon_a + \epsilon_b -
                       \epsilon_p - \epsilon_q} \right ].
  \label{algebra_e2}
\end{equation}
In the above expression, the first and second terms on the right hand side
are the direct and exchange. Though the expression is fairly straight
forward to derive, we have given explicitly for  easy reference while 
analysing the results.


\subsection{Third-Order Correlation Energy}

 The diagrammatic representation of the second order wave operator 
$\Omega^{(2)}$ consists of single, double, triple and quadruple excitations. 
The singles are non-zero starting from the second order when Dirac-Fock 
orbitals are used. And, the triples and quadruples begin to contribute from 
this order. The triples consist of connected diagrams, whereas all the 
quadruples are disconnected. The third order correlation energy is
\begin{equation}
  E^{(3)}_{\rm corr} = \langle\Phi_0|(V_1 + V_2)(\Omega^{(2)}_1
            + \Omega^{(2)}_2 + \Omega^{(2)}_3 + \Omega^{(2)}_4)|\Phi_0\rangle.
\end{equation}
The triple and quadruple excitations do not contribute as $V$ at the most
can contract with double excitations. For the same reason mentioned earlier, 
in second order,  $V_1$ also does not contribute. Then the third order 
correlation energy is simplified to
\begin{equation}
  E^{(3)}_{\rm corr} = \langle\Phi_0|V_2\Omega^{(2)}_2)|\Phi_0\rangle.
  \label{corr_e3}
\end{equation}
This is similar in form to the second order correlation energy. In general,
the $n^{\rm th}$ order correlation energy has non-zero contribution from the
term $V_2\Omega^{(n-1)}_2$ only.  It must be mentioned that the 
connected triples begin to contribute from the fourth order energy. This
is utilized in perturbative inclusion of triples, in later sections of the 
paper, while discussing coupled-cluster calculations.


\section{Coupled-Cluster Theory}

The coupled-cluster theory is a non-perturbative many-body theory and 
considered as one of the best. A recent review \cite{bartlett-07} provides an
excellent overview of recent developments and different variations.
In the context of diagrammatic analysis of many-body perturbation theory, 
coupled-cluster theory is equivalent to a selective evaluation of the 
connected diagrams to all orders. Then casting the disconnected but linked 
diagrams as products of connected diagrams. In coupled-cluster theory, for a 
closed-shell atom, the exact ground state is
\begin{equation}
  |\Psi_0\rangle = e^{T^{(0)}}|\Phi_0\rangle,
  \label{ccwave}
\end{equation}
where $T^{(0)}$ is the cluster operator. The superscript is a tag to identify
cluster operators arising from different perturbations. For the case of $N$ 
electron atoms, the cluster operator is
\begin{equation}
  T^{(0)}= \sum_{i=1}^N T^{(0)}_i.
  \label{t}
\end{equation}
%
%
\begin{figure}
  \includegraphics[width = 5.0 cm]{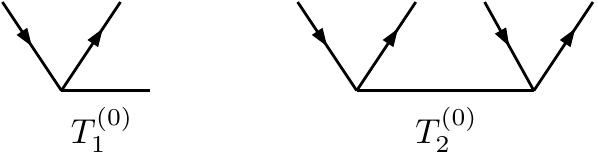}
  \caption{Diagrammatic representation of unperturbed single and double
           cluster operators.}
\end{figure}
In closed shell atoms, the single and doubles provide a good approximation of 
the exact ground state. Then, the cluster operator 
$T^{(0)}=T_1^{(0)} + T_2^{(0)}$ and is referred to as the coupled-cluster 
single and doubles (CCSD). The cluster operators in the  second quantized 
notations are
\begin{eqnarray}
  T^{(0)}_1 & = &\sum_{a, p}t_a^p a_p^{\dagger}a_a, \\ 
  T^{(0)}_2 & = &\frac{1}{2!}\sum_{a, b, p, q}t_{ab}^{pq}.
  a_p^{\dagger}a_q^{\dagger}a_ba_a.
\end{eqnarray}
Here, $t_a^p$ and $t_{ab}^{pq}$ are the  single and double cluster 
amplitudes respectively.  Subtracting $\langle \Phi_0|H|\Phi_0\rangle$ from both sides of
Eq.(\ref{dcpsi}) and using the normal form of an operator, 
$ O_N = O - \langle \Phi_0|O|\Phi_0\rangle$, we get
\begin{equation}
  H_{\rm N}|\Psi_0\rangle = \Delta E|\Psi_0\rangle,
  \label{deltaE}
\end{equation}
where $\Delta E = E - \langle \Phi_0|H|\Phi_0\rangle$, as defined earlier, is 
the correlation energy. Operating with $e^{-T^{(0)}}$ and projecting the above 
equation on excited states we get the cluster amplitude equations  
\begin{eqnarray}
  \langle\Phi^p_a|\overline{H}_{\rm N}|\Phi_0\rangle = 0,
  \label{cc1eq}    \\
  \langle\Phi^{pq}_{ab}|\overline H_{\rm N}|\Phi_0\rangle = 0,
  \label{cc2eq}
\end{eqnarray}
where $\overline{H}_{\rm N}=e^{-T^{(0)}}H_{\rm N}e^{T^{(0)}} $ is the 
similarity transformed or dressed Hamiltonian. Following Wick's theorem and 
structure of $H_{\rm N}$, in general 
\begin{eqnarray}
  \overline{H}_{\rm N}=&&H_{\rm N}+\{\contraction{}{H}{_{\rm N}}{T}
   H_{\rm N}T^{(0)}\} +
  \frac{1}{2!}\{\contraction{}{H}{_{\rm N}}{T}
  \contraction[1.5ex]{}{H}{_{\rm N}T^{(0)}}{T} H_{\rm N}T^{(0)}T^{(0)}\} +
  \nonumber \\
  &&\frac{1}{3!}\{\contraction{}{H}{_{\rm N}}{T}
  \contraction[1.5ex]{}{H}{_{\rm N}T^{(0)}}{T}
  \contraction[2ex]{}{H}{_{\rm N}T^{(0)}T^{(0)}}{T}
  H_{\rm N}T^{(0)}T^{(0)}T^{(0)}\}  
  + \frac{1}{4!}\{\contraction{}{H}{_{\rm N}}{T}
  \contraction[1.5ex]{}{H}{_{\rm N}T^{(0)}}{T}
  \contraction[2ex]{}{H}{_{\rm N}T^{(0)}T^{(0)}}{T}
  \contraction[2.5ex]{}{H}{_{\rm N}T^{(0)}T^{(0)}T^{(0)}}{T}
  H_{\rm N}T^{(0)}T^{(0)}T^{(0)}T^{(0)}\},
  \label{Hnbarcont}
\end{eqnarray}
Here $\contraction{}{A}{\ldots}{B}A\ldots B$ denote contraction between two
operators $A$ and $B$. The single and double cluster amplitudes are solutions 
of Eq.(\ref{cc1eq}) and (\ref{cc2eq}) respectively. These are set of coupled 
nonlinear equations and iterative methods are the ideal choice to solve
these equations.
%
%
\begin{center}
\begin{figure}
  \includegraphics[width = 8.0cm]{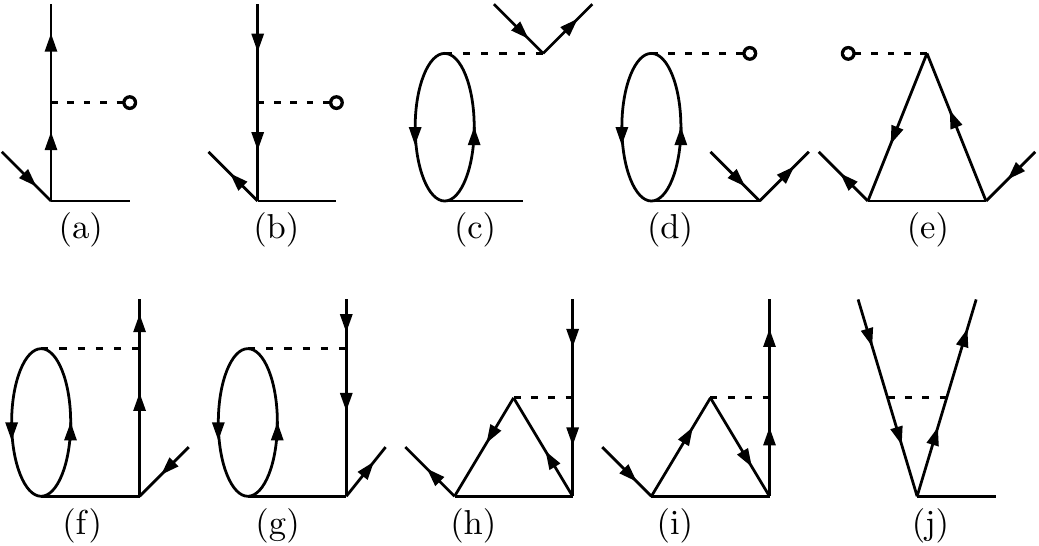}
  \caption{Diagrams which contribute to the singles, unperturbed cluster
           operator ($T^{(0)}_1 $), in the linearised coupled-cluster.}
\end{figure}
\end{center}
%
%
\begin{center}
\begin{figure}
  \includegraphics[width = 8.0cm]{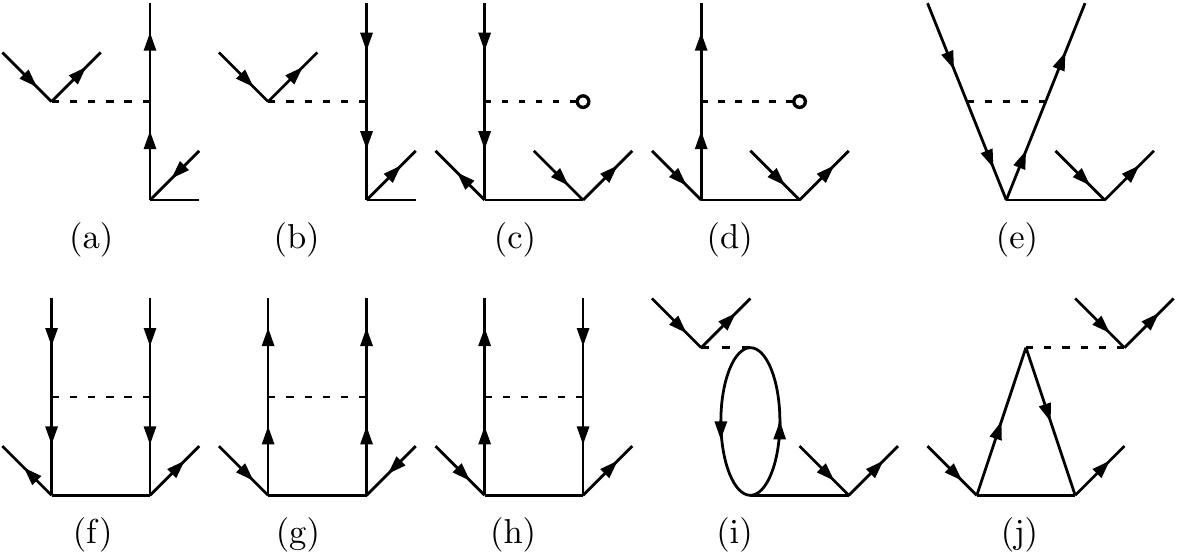}
  \caption{Diagrams which contribute to the doubles, unperturbed cluster
           operator ($T^{(0)}_2 $), in the linearised coupled-cluster.}
\end{figure}
\end{center}
%
%


\subsection{Linearized Coupled-Cluster}

The nonlinearity in the cluster amplitude equation arises from the two and 
higher contractions in the dressed Hamiltonian.  An approximation often used
as a starting point of coupled-cluster calculations is to retain only
the first two terms in $\overline{H}_N$, then
\begin{equation}
  \overline{H}_{\rm N} = H_{\rm N}+\{\contraction{}{H}{_{\rm N}}{T}H_{\rm N} 
  T^{(0)}\} .
\end{equation}
The cluster equations are then a pair of linear equations 
\begin{eqnarray}
  \langle\Phi^p_a|H_{\rm N}+\{\contraction{}{H}{_{\rm N}}{T}H_{\rm N} T^{(0)}\}
   |\Phi_0\rangle = 0,  \\
  \langle\Phi^{pq}_{ab}|H_{\rm N}+\{\contraction{}{H}{_{\rm N}}{T}H_{\rm N}
  T^{(0)}\}|\Phi_0\rangle = 0. 
\end{eqnarray}
In the CCSD approximation $ T^{(0)} = T^{(0)}_1 + T^{(0)}_2$, these equations
are then
\begin{eqnarray}
  \langle\Phi^p_a|\{\contraction{}{H}{_{\rm N}}{T}H_{\rm N} T^{(0)}_1\} +
   \{\contraction{}{H}{_{\rm N}}{T}H_{\rm N} T^{(0)}_2\}|\Phi_0\rangle =
  -\langle\Phi^p_a|H_{\rm N}|\Phi_0\rangle\nonumber \\
  \langle\Phi^{pq}_{ab}|\{\contraction{}{H}{_{\rm N}}{T}H_{\rm N} T^{(0)}_1\} +
  \{\contraction{}{H}{_{\rm N}}{T}H_{\rm N} T^{(0)}_2\}|\Phi_0\rangle =
  -\langle\Phi^{pq}_{ab}|H_{\rm N}
  |\Phi_0\rangle. \nonumber \\
\end{eqnarray}
These are the linearized coupled-cluster equations of single and double cluster
amplitudes. This can be combined as the matrix equation
\begin{eqnarray}
  \left(\begin{array}{cc}
                H_{11} & H_{12} \\
                H_{21} & H_{22}
        \end{array}\right)
  \left(\begin{array}{c}
                t_1 \\
                t_2
        \end{array}\right) = -
  \left(\begin{array}{c}
                H_{10}\\
                H_{20}
        \end{array}\right),
\end{eqnarray}
where $H_{11} =\langle\Phi^p_a|H_{\rm N}|\Phi^s_b\rangle $,
$H_{12} =\langle\Phi^p_a|H_{\rm N}|\Phi^{st}_{bc}\rangle$ and so on.
The equations are set of coupled linear equations and solved using 
standard or specialized linear algebra solvers. In the literature several
authors refer to linearized coupled-cluster as all-order method. A 
description of the all-order method  and applications  are given in 
ref \cite{safronova-07}. In a recent work, the authors report the 
combination of all-order method and configuration interaction
\cite{safronova-09}.


\subsection{Correlation Energy and approximate triples \label{corr_en}}

\begin{center}
\begin{figure}[h]
  \includegraphics[width = 8.0cm]{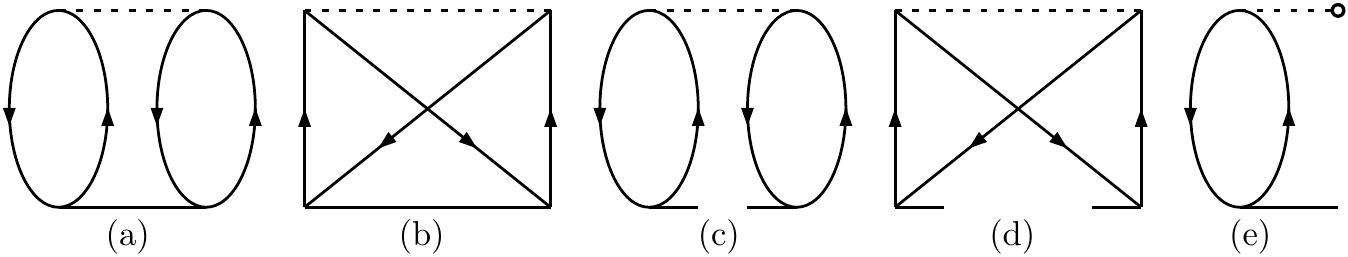}
  \caption{Coupled-cluster correlation energy diagrams. The diagram (e) is 
           equal to zero when Dirac-Fock orbitals are used.}
  \label{cccorr}
\end{figure}
\end{center}

 From Eq.(\ref{deltaE}) the ground state correlation energy, in coupled-cluster
theory, is the ground state expectation value of $\overline{H}_N$. That is 
\begin{equation}
  \Delta E = \langle\Phi_0|\overline{H}_{\rm N}|\Phi_0\rangle.
\end{equation}
The diagrams arising from the above expression are
shown  in the Fig.\ref{cccorr}. The dominant contributions are from the 
diagrams (a) and (b), which is natural as the doubles cluster amplitudes are 
larger in value than the singles. Diagram (e) does not contribute to the 
correlation energy when Dirac-Fock orbitals are used.
%
%

\begin{center}
\begin{figure}
  \includegraphics[width = 8cm]{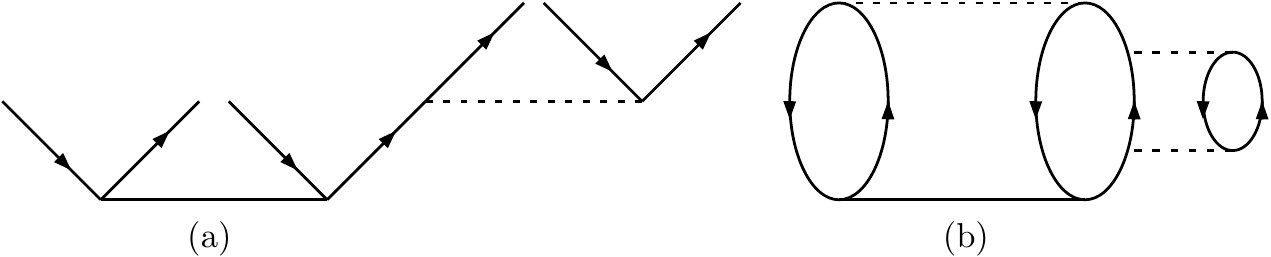}
  \caption{Diagrams of approximate triples calculated perturbatively:(a) 
           approximate triples cluster operator and (b) correlation energy
           arising from approximate triples.}
  \label{triples}
\end{figure}
\end{center}

 To go beyond the CCSD approximation, we incorporate selected correlation 
energy diagrams arising from approximate triples. The approximate triples are 
perturbative contraction of $V_2$ with the $T^{(0)}$ cluster amplitudes
\cite{krishnan-89,porsev-06}. Example diagrams of the approximate triples and 
correlation energies are shown in Fig.\ref{triples}. There are two categories 
of triples, first is $V_2$ contracted with $T^{(0)}$ through a hole line, and 
second contraction through a particle line (Fig.\ref{triples}a). To calculate 
the correlation energy from the triples, these are contracted perturbatively 
with $V_2$ and reduced to a double excitation diagram. Then the correlation 
energy is obtained after another contraction with $V_2$. These two contractions
generate several diagrams. The triples correlation energy diagrams 
are separated into three categories based on the number of internal lines. 
These are: two particle and two hole internal lines (2p-2h), three 
particle one hole internal lines (3p-1h), and one particle three hole internal
lines (1p-3h). In the present calculations eight diagrams from the first 
category and two each from other remaining two categories are considered. 
%
%
%
%


\section{Perturbed Coupled-Cluster}

  The atomic properties of interest are, in general,  
associated with additional interactions. The interaction are either internal 
like hyperfine interaction or external like static electric field. These are 
treated as perturbations which modify the wave function and energy of the atom. 
This section briefly describes a method to incorporate an additional 
perturbation within the frame work of relativistic coupled-cluster. The scheme 
is referred to as perturbed coupled-cluster theory. It has been tried and 
tested in precision atomic properties and structure calculations. In the 
presence of a perturbation $H_1$, the eigen value equation is
\begin{equation}
  \left ( H^{\rm DC} + \lambda H_1\right )|\widetilde\Psi_0\rangle=
      \widetilde {E}|\widetilde\Psi_0\rangle.
  \label{pdcpsi}
\end{equation}
Here $\vert\widetilde\Psi_0\rangle$ is the perturbed wave function,
$\widetilde {E}$ is the corresponding eigenvalue and $\lambda$ is the 
perturbation parameter. The perturbed wave function is the sum of the 
unperturbed wave function and a correction $|\overline{\Psi}^1_0 \rangle$ 
arising from $H_1$. That is
\begin{equation} 
  |\widetilde{\Psi}_0 \rangle = |\Psi_0 \rangle + 
      \lambda|\overline{\Psi}^1_0 \rangle.
\end{equation} 
%
%
\begin{center}
\begin{figure}
  \includegraphics[width = 5.0 cm]{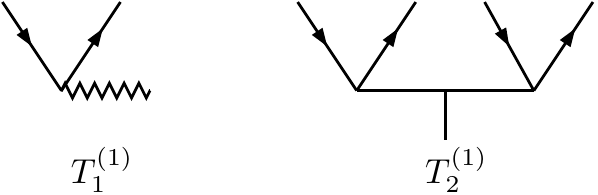}
  \caption{Diagrams of single and double perturbed cluster operators.}
  \label{pt1t2}
\end{figure}
\end{center}
Following the earlier description of coupled-cluster wave function, the 
perturbed wave function is 
\begin{equation}
  |\widetilde\Psi_0\rangle = e^{T^{(0)} + \lambda T^{(1)}}|\Phi_0\rangle .
  \label{pccwave1}
\end{equation}
The cluster operators $T^{(0)}$, as defined earlier, incorporate the effects
of residual Coulomb interaction. For clarity these are referred as unperturbed 
cluster operator. The $T^{(1)}$ cluster operators arise from $H_1$ and 
are referred to as the perturbed cluster operators. It acts on the reference
state $|\Phi_0\rangle$ to generate the correction. Consider the perturbation
expansion to first order in $\lambda$,  we get
\begin{equation}
  |\widetilde\Psi_0\rangle = e^{T^{(0)}} (1 + \lambda T^{(1)})|\Phi_0\rangle.
  \label{pccwave2}
\end{equation}
To derive the cluster equations use this in Eq.(\ref{pdcpsi}), then operate 
with $e^{-T^{(0)}}$ and project on excited states. We get the equations for 
singles and doubles perturbed cluster amplitudes
\begin{eqnarray}
  \langle\Phi^p_a|\{\contraction{}{H}{_{\rm N}}{T}\overline{H}_{\rm N} 
    T^{(1)}\} |\Phi_0\rangle = 
  -\langle\Phi^p_a|\bar H_1|\Phi_0\rangle, \\
  \langle\Phi^{pq}_{ab}|\{\contraction{}{H}{_{\rm N}}{T}
  \overline{H}_{\rm N}T^{(1)}\} |\Phi_0\rangle = 
  -\langle\Phi^{pq}_{ab}|\overline{H}_1|\Phi_0\rangle.
\end{eqnarray}
The dressed Hamiltonian $\overline{H}_{\rm N}$ is same as in 
Eq.(\ref{Hnbarcont}). Like in linearized coupled-cluster, these form a set of
linear algebraic equations.


\subsection{Approximate triples}

\begin{center}
\begin{figure}
  \includegraphics[width = 8.0cm]{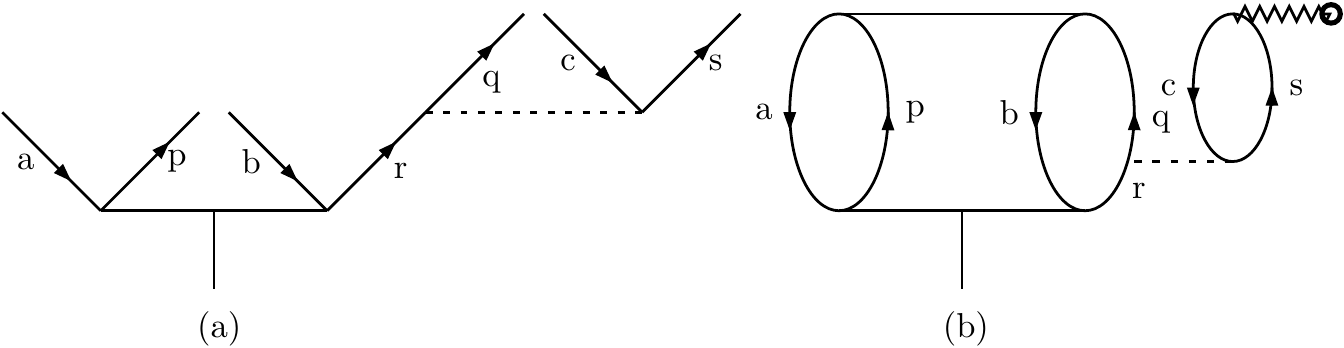}
  \caption{Diagrams of approximate triples calculated perturbatively:(a) 
           (a) Representation of approximate perturbed triples.
           (b) Contribution of approximate perturbed triples to the 
               dipole polarizability.}
  \label{t3dp}
\end{figure}
\end{center}

 Like in $T^{(0)}$, a perturbed triple cluster Fig.\ref{t3dp}(a) is 
a perturbative contraction between $V_2$ and $T^{(1)}_2$. As in the case of 
unperturbed approximate triples discussed earlier, there are two types of 
diagrams in the present case as well. One arises from particle line 
contraction and the other from hole line contraction between $V_2$ and
$T^{(1)}$ diagrams.  In this work we implement approximate triples while
calculating properties. In particular, to calculate dipole polarizability and
an example diagram is shown in Fig. \ref{t3dp}(b). The algebraic expression
of this diagram is
\begin{equation}
  \sum_{a,b,c,p,q,r,s}
  \frac{\langle ab|{T^{(0)}_2}^\dagger|pq\rangle \langle c|\bm{d}|s\rangle
        \langle qs|V_2|rc\rangle \langle pr|T^{(1)}_2|ab\rangle}
  {\epsilon_a+\epsilon_b+\epsilon_c
  -\epsilon_p-\epsilon_q-\epsilon_s},
\end{equation}
here, $\bm{d}$ is the dipole operator. In total there are twentyfour properties 
diagrams arising from the perturbative triples and we include all of these
diagrams in the calculations.


\subsection{Dipole Polarizability}
When an atom is placed in an external electric field $\cal \bm{E}$, the 
charge distribution of electron cloud is distorted and an electric dipole 
moment $\bm{D}_{\rm ind}$ is induced. The dipole polarizability of the atom 
$\alpha$ is then the ratio of the induced dipole moment to the applied 
electric field, that is
\begin{equation}
  \bm{D}_{\rm ind} = \alpha \cal \bm{E}.
  \label{dind1}
\end{equation}
By definition, the dipole polarizability of the ground state is 
\begin{equation}
  \alpha = -2 \sum_I \frac{|\langle\Psi_0|\bm{D}|\Psi_I \rangle|^2}{E_0-E_I},
  \label{alpha}
\end{equation}
where $|\Psi_I \rangle$ are intermediate atomic states. These are opposite in 
parity to the ground state $|\Psi_0\rangle$. The expression of $\alpha$ 
can be rewritten as
\begin{equation}
  \alpha = -2 \langle\Psi_0|\bm{D}|\overline{\Psi}_0^1\rangle. 
  \label{alpha2}
\end{equation}
Here 
$|\overline{\Psi}_0^1\rangle = 
\sum_I (|\Psi_I\rangle\langle\Psi_I|\bm{D}|\Psi_0\rangle)/(E_0-E_I) $, which
follows from the first order time independent perturbation theory.
The perturbation Hamiltonian is $H_1 = -\bm{D}.\cal \bm{E}$ and 
external field $\cal \bm{E}$ is the perturbation parameter. One short coming
of calculating $\alpha$ from Eq.(\ref{alpha}) is, for practical reasons, the 
summation over $I$  is limited to the most dominant intermediate states.
However, the summation is avoided altogether when the perturbed 
coupled-cluster wave functions are used in the calculations. From  
Eq.(\ref{pccwave1}) the perturbed wave function
\begin{equation}
  |\overline{\Psi}_0^1\rangle = e^{T^{(0)}}T^{(1)}|\Phi_0\rangle .
\end{equation}
 In a more compact form, the dipole polarizability in terms of the 
perturbed coupled-cluster wave function is 
\begin{equation}
  \alpha =\langle \widetilde\Psi_0|\bm{D} |\widetilde\Psi_0 \rangle.
\end{equation}
After simplification, using the perturbed wave function in 
Eq.(\ref{pccwave1}), we get
\begin{equation}
  \alpha =\langle \overline\Psi_0^1|\bm{D} |\Psi_0 \rangle +
          \langle \Psi_0|\bm{D} |\overline{\Psi}_0^1 \rangle.
\end{equation}
The correction $ |\overline{\Psi}_0^1\rangle$, as described earlier, is 
opposite in parity to $|\Phi_0\rangle$. Hence the matrix elements  
$\langle\Psi_0|\bm{D} |\Psi_0 \rangle$
and $\langle\overline\Psi_0^1|\bm{D} |\overline{\Psi}_0^1\rangle$ are zero.
As $D$ is hermitian, the two terms on the left hand side are identical and 
the above expression is same as Eq.(\ref{alpha2}). Considering the leading 
terms
\begin{equation}
  \alpha =\langle\Phi_0|{T^{(1)}}^\dagger \overline{D}^{(0)} + 
          \overline{D}^{(0)} T^{(1)}|\Phi_0\rangle. 
  \label{alpha_cc}
\end{equation}
Here, the operator
$\overline{D}^{(0)} = e^{{T^{(0)}}^\dagger} D e^{T^{(0)}}$ is the unitary 
transformed electric dipole operator. It is explicitly evident that the
dipole polarizability, in terms of perturbed cluster operator, does not 
have a sum over states. In this scheme, contributions from all intermediate 
states within the chosen configuration space are included. For precision 
calculations, this is a very important advantage.
%
%
%
%

\section{Results}

  In order to get reliable results, in atomic structure and properties 
calculations, one prerequisite is good quality orbital basis set.  
In all calculations described in the paper, we employ GTOs as orbital 
functions. In particular, we use even tempered 
basis in which the parameters $\alpha_0$ and $\beta$,
in Eq. (\ref{param_gto}), are different for each symmetries. We use the 
basis parameters of Tatewaki and Watanabe \cite{tatewaki-04} as starting 
values and optimized further to obtain $E_{\rm DC}^{(0)}$ (ground state 
Dirac-Fock energy) and $\epsilon_a$ (single particle energies of core 
orbitals) in agreement with the numerical results. The numerical results are 
obtained from  the GRASP92 \cite{parpia-96} code. In order to obtain converged 
$E^{(2)}_{\rm corr}$, we consider orbital basis set consisting of all the 
core orbitals and virtual orbitals up to 10,000--11,000 in single particle 
energies.

  The working equations of coupled-cluster theory are coupled nonlinear
equations. Solving these equations is a computational challenge. The number
of unknowns, cluster amplitudes, are in the order of millions. In addition,
implementing fast and efficient algorithms demand huge memory to tabulate
and store two-electron integrals. This is essential as the two-electron
integrals are needed repeatedly and are compute intensive. For the larger basis 
sets in the present work, the number of two-electron integrals is more than
$2\times 10^8$. In order to utilize memory efficiently, we have developed a
scheme which parallelize the tabulation and storage of two-electron 
integrals. To improve performance further, we also tabulate and store $6j$ 
symbols. This is desirable as the angular part of the perturbed cluster 
diagrams involve large number of $6j$ symbols. Quantitatively, we observe a 
performance gain of ~30\% or more with the $6j$ symbols tabulation. We shall
describe and discuss the various computational schemes developed in a 
future publication.

  The unperturbed and perturbed cluster amplitude equations are solved
iteratively using Jacobi method. We chose the method as it is relatively
simple to parallelize. One drawback of the method is, it converges slowly.
To obtain faster convergence, we employ direct inversion in iterated subspace
(DIIS) \cite{pulay-80} convergence acceleration.

\subsection{Second-Order Correlation Energy}
\begin{table*}[t]
\caption{The SCF $E^{(0)}_{DC}$, the second-order correlation 
         $E^{(2)}_{\rm corr}$ and the total energies $E$ of  Ne, Ar, Kr and 
         Xe. All the values listed are in atomic units (Hartrees).}
\begin{ruledtabular}
\begin{tabular}{ccccccccc}
Atom&   Z & \mbox{Atomic mass} &\multicolumn{3}{c}{\mbox{This work}}&
    \multicolumn{3}{c} {\mbox{Other work}}    \\ \hline
    &     &       &\mbox{$E^{(0)}_{DC}$}&\mbox{$E^{(2)}_{\rm corr}$}  & $E$
                  &\mbox{$E^{(0)}_{DC}$}&\mbox{$E^{(2)}_{\rm corr}$}  & $E$ \\
\hline
 Ne & 10 & 20.18  & $-128.6932$  & $-0.3830$                  & $-129.0762$
                  & $-128.6919$  & $-0.3834$\footnotemark[1]  & $-129.0753$ \\
    &     &       &              &                            &
                  &              & $-0.3836$\footnotemark[2]  &             \\
    &     &       &              &                            &
                  &              & $-0.3822$\footnotemark[3]  &             \\
    &     &       &              &                            &
                  &              & $-0.3697$\footnotemark[4]  &             \\
    &     &       &              &                            &
                  &              & $-0.3804$\footnotemark[5]  &             \\
 Ar & 18 & 39.95  & $-528.6882$  & $-0.6938$                  & $-529.3820$
                  & $-528.6838$  & $-0.6981$\footnotemark[1]  & $-529.3819$ \\
    &     &       &              &                            &
                  &              & $-0.6822$\footnotemark[5]  &             \\
    &     &       &              &                            &
                  &              & $-0.685$\footnotemark[6]   &             \\
    &     &       &              &                            &
                  &              & $-0.790$\footnotemark[7]   &             \\
 Kr & 36 & 83.80  & $-2788.8659$ & $-1.8426$                  & $-2790.7085$
                  & $-2788.8615$ & $-1.8468$\footnotemark[1]  & $-2790.7083$ \\
 Xe & 54 & 131.29 & $-7446.8887$ & $-2.9767$                  & $-7449.8654$
                  & $-7446.8880$ & $-2.9587$\footnotemark[1]  & $-7449.8467$ \\
\end{tabular}
\end{ruledtabular}
  \footnotetext[1]{Reference\cite{Ishikawa-94}.}
  \footnotetext[2]{Reference\cite{lindgren-80}.}
  \footnotetext[3]{Reference\cite{Nesbet-68}.}
  \footnotetext[4]{Reference\cite{Sasaki-74}.}
  \footnotetext[5]{Reference\cite{Ishikawa-90}.}
  \footnotetext[6]{Reference\cite{Cooper-73}.}
  \footnotetext[7]{Reference\cite{Clementi-65}.}
  \label{engy_total}
\end{table*}
  The SCF energy $E^{(0)}_{\rm DC}$, second-order correlation energy 
$E^{(2)}_{\rm corr}$ and the total energy $E$ from our calculations are 
listed in Table.\ref{engy_total}. For comparison the results of previous 
calculations are also listed. It is evident that our second-order correlation 
energy and total energy, sum of the SCF and second order correlation energy, 
are in agreement with the results of Ishikawa et al \cite{Ishikawa-94} for 
all the atoms studied. Among the previous works, we select the results of 
Ishikawa et al \cite{Ishikawa-94} for detailed comparison as their 
calculations are relativistic. The other results listed in the table are from 
non-relativistic calculations. For all the atoms, our SCF energy 
$E_{\rm DC}^0$ are lower and there are no discernible trends, as a function
of nuclear charge, in the difference. Interestingly, except for Xe, our second 
order correlation energies are higher. This compensates the lower 
$E^{(0)}_{\rm DC}$ and subsequently, the total energies $E$ of the two 
calculations are in excellent agreement. The lack of 
trend indicates the choice of the basis set could be the source of the
observed differences of $E^{(0)}_{\rm DC}$ and $E^{(2)}_{\rm corr}$ between the 
two calculations.
%
%
\begin{figure}[h]
\begin{center}
  \includegraphics[width=8.0cm]{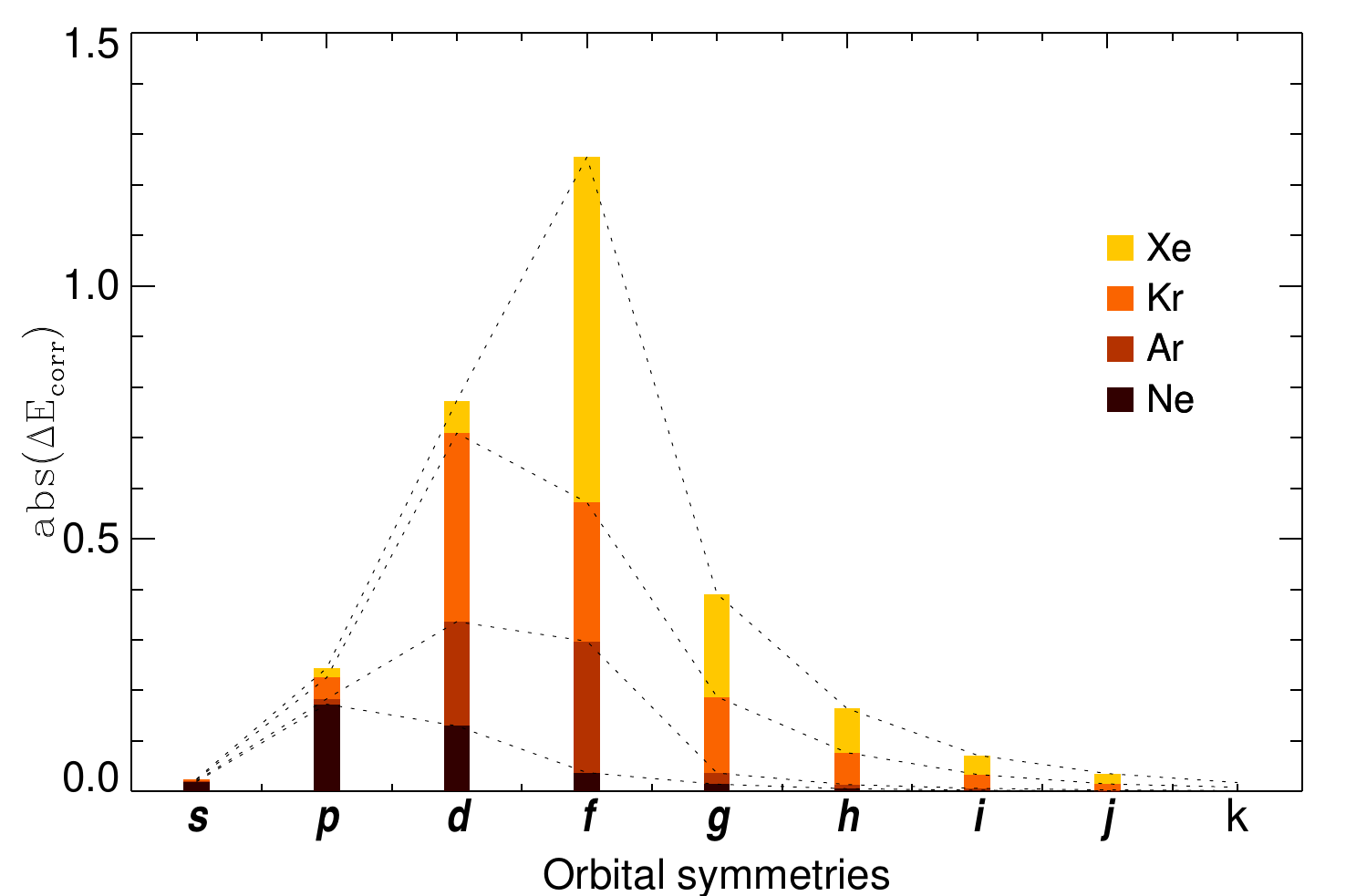}
  \caption{The second-order correlation energy when orbitals upto a particular
           symmetry are included in the virtual space.}
  \label{corr_smy}
\end{center}
\end{figure}
%
%
\begin{table}[h]
\caption{Cumulative second-order correlation energy when orbitals
         upto a particular symmetry are included in the virtual space. All the
         values are in atomic units.}
\begin{ruledtabular}
\begin{tabular}{ccccc}
 Symmetry &     Ne    &     Ar    &     Kr    &    Xe    \\ \hline
    $s$   &  -0.0194  &  -0.0210  &  -0.0236  &  -0.0247 \\
    $p$   &  -0.1920  &  -0.2043  &  -0.2479  &  -0.2687 \\
    $d$   &  -0.3216  &  -0.5401  &  -0.9512  &  -1.0419 \\
    $f$   &  -0.3589  &  -0.6330  &  -1.5213  &  -2.2972 \\
    $g$   &  -0.3732  &  -0.6695  &  -1.7077  &  -2.6879 \\
    $h$   &  -0.3786  &  -0.6830  &  -1.7843  &  -2.8520 \\
    $i$   &  -0.3811  &  -0.6891  &  -1.8179  &  -2.9238 \\
    $j$   &  -0.3823  &  -0.6921  &  -1.8343  &  -2.9591 \\
    $k$   &  -0.3830  &  -0.6938  &  -1.8426  &  -2.9767 \\
\end{tabular}
\end{ruledtabular}
  \label{engy_sym}
\end{table}
   The Table.\ref{engy_sym} lists the cumulative contributions from various 
symmetries to $E^{(2)}_{\rm corr}$. Among the previous works, Lindgren and 
collaborators \cite{lindgren-80} provide cumulative $E^{(2)}_{\rm corr}$ for 
Ne up to $i$ symmetry. Their converged result is -0.3836,  this compares well 
with our result of -0.3811 calculated with orbitals up to $i$ symmetry. However,
in our calculation, we get converged results of -0.3830 after including $j$ 
and $k$ symmetry orbitals. For Ne, $E^{(2)}_{\rm corr}$ is considered 
converged when the change with additional symmetry is below millihartree. 
However, for Ar, Kr and Xe orbitals of $l$ and higher symmetries are essential 
to obtain $E^{(2)}_{\rm corr}$ converged up to millihartree.  Since the magnitude 
of $E^{(2)}_{\rm corr}$ increases with $Z$, a representative measure of 
convergence is the percentage change. The contribution from $k$ 
symmetry to $E^{(2)}_{\rm corr}$ are -0.0017, -0.0083 and -0.0176 for Ar, Kr 
and Xe respectively. These are larger than that of Ne, which is -0.0007. 
However, these correspond to 0.24\%, 0.45\% and 0.59\% for Ar, Kr and Xe
respectively, these compare very well with that of Ne 0.20\%.
For Ar there is a variation in the previous values of $E^{(2)}_{\rm corr}$,
these range from the lowest value of Clementi \cite{Clementi-65} 
-0.790 to that of Ishikawa \cite{Ishikawa-94} -0.6981. Our value of -0.6938 is 
closer to that of Ishikawa. 

  There is a  pattern  in the change of the correlation energy with symmetry 
wise augmentation of the virtual orbital set. There is an initial increase,
reaches a maximum and then decreases. The maximum change occurs with the 
addition of $p$, $d$, $d$ and $f$ symmetry for Ne, Ar, Kr and Xe respectively. 
The pattern is evident in Fig.\ref{corr_smy}, which plots the change in 
$E^{(2)}_{\rm corr}$ with symmetry wise augmentation of the virtual space. 
The pattern arise from the distribution of the contributions from each of 
the core orbitals. From Eq.(\ref{algebra_e2}), depending on the core orbital 
combination $ab$, there are two types of correlation effects. These are
inter and intra core shell correlations corresponding to $a=b$ and $a\neq b$
respectively. Among the various combinations, the $2p_{3/2}2p_{3/2}$,
$3p_{3/2}3p_{3/2}$, $3d_{5/2}3d_{5/2}$ and $4d_{5/2}4d_{5/2}$ core orbital
pairings have leading contributions in Ne, Ar, Kr and Xe respectively. Here
for Ne and Ar the leading pairs correspond to the valence shell but it is 
the last $d$ shell for  Kr and Xe. This correlates with the pattern observed 
in the symmetry wise augmentation.


\subsection{Third-Order Correlation Energy}
   We calculate the third order correlation energy $E^{(3)}_{\rm corr}$ from 
the linearized CCSD equations. This is possible when the first 
order MBPT wave operator $\Omega^{(1)}$  is chosen as the initial guess and 
iterate the linearized coupled-cluster equations once. The two-body 
wave operator so obtained is $\Omega_2^{(2)}$ and from Eq.(\ref{corr_e3})
$E^{(3)}_{\rm corr}=\langle\Phi_0|V_2\Omega_2^{(2)}|\Phi_0\rangle $. This
approach, however, is not applicable beyond third order. The reason is, 
starting from the fourth order correlation energy the triples contribute 
to $\Delta E $ and triples are not part of the linearized CCSD equations.
The results of $E^{(3)}_{\rm corr}$, obtained from our calculations, are 
listed in Table.\ref{engy_third}.  For comparison, results from previous works
are also listed. For Ne, Jankowski and Malinowski \cite{Jankowski} reported
a value of 0.0024. Their calculations were done with a limited basis set and 
hence, could leave out less significant contributions. The results of 
Lindgren  and collaborators \cite{lindgren-80} 0.0035 is perhaps more accurate
and reliable on account of larger basis set. They include virtuals up to $i$ 
symmetry and then extrapolate. Similarly, in our calculations we include 
virtual orbitals up to $i$ symmetry, then extrapolate up to $k$ symmetry based
on $E^{(2)}_{\rm corr}$ results. We obtain 0.0019, which is in better agreement 
with the result of Jankowski and Malinowski\cite{Jankowski}. As expected, 
$E^{(3)}_{\rm corr}$ increases with $Z$ and to our knowledge, our results of 
Ar, Kr and Xe are the first reported calculations in literature. Interestingly,
$E^{(3)}_{\rm corr}$ is positive for Ne, Kr and Xe but it is negative for Ar. 
\begin{table}[h]
\caption{Third-order correlation energy in atomic units.}
\begin{ruledtabular}
\begin{tabular}{ccc}
Atom &\multicolumn{2}{c}{$E_3$}                    \\
\hline
  & This work & Other work                         \\
\hline
 Ne   & 0.0019           & 0.0035\footnotemark[1]  \\
      &                  & 0.0024\footnotemark[2]  \\
 Ar   & -0.0127           &  - \\
      &                  &  - \\
 Kr   & 0.0789           &  - \\
      &                  &  - \\
 Xe   & 0.1526           &  - \\
      &                  &  - \\
\end{tabular}
\end{ruledtabular}
\footnotetext[1]{Reference\cite{lindgren-80}.}
\footnotetext[2]{Reference\cite{Jankowski}.}
  \label{engy_third}
\end{table}
%


\subsection{Coupled-Cluster correlation energy}
    The MBPT correlation energies $E_{\rm corr}^{(i)}$ converges with 
relatively large basis set. For example, the $E_{\rm corr}^{(2)}$ of Ne
converge when virtual orbitals up to $k$ symmetry are included in the 
calculations. This correspond to a total of 224 virtual orbitals. 
Similar or larger number of virtual orbitals are required to obtain
converged $E_{\rm corr}^{(2)}$ of Ar, Kr and Xe as well. However, it is not
practical to have such large basis sets in relativistic coupled-cluster 
calculations. The $n_v^4n_c^3$, where $n_v$ and $n_c$ are the number of the
virtual and core orbitals respectively, scaling of arithmetic operations in  
CCSD render computations with large $n_v$ beyond the scope of detailed 
studies. Hence, in the CCSD calculations, the size of the virtual orbital set 
is reduced to a manageable level and restrict up to the $h$ symmetry. To choose 
the optimal set, after considering the most dominant ones, the virtual 
orbitals are augmented in layers. Where one layer consists of one virtual 
orbital each from all the symmetries considered.

   The CCSD correlation energies for two basis sets  are listed in 
Table.\ref{engy_cct}. The first is with a basis set considered optimal
and manageable size for CCSD calculations after a series of calculations. 
Then the next is with an additional layer of virtual orbitals. The  change
in the linearized CCSD $\Delta E$ with the additional layer of virtual 
orbitals are 0.2\%, 1.7\%, 6.0\% and 5.0\% for Ne, Ar, Kr and Xe respectively.
Changes of similar order are observed in the corresponding $\Delta E$ of the 
non-linear CCSD calculations. It must be mentioned that, though the difference 
in  $\Delta E$ is small, the computational cost of non-linear CCSD is much  
higher than the linearized CCSD calculations. The percentage changes indicate
the  basis size of Kr and Xe are not large enough. The orbital basis of Xe,
with the additional layer, consists of 17 core and 129 virtual
orbitals. This translates to $\sim 5.0 \times 10^{6}$ cluster amplitudes, 
which follows from the $n_v^2n_c^2$ scaling of the number of cluster 
amplitudes. At this stage, the computational efforts and
costs far out weight the gain in accuracy. To account for the correlation
energy from the other virtual orbitals, not included in the CCSD calculations,
we resort to the second order correlation energy. For this we calculate 
$E_{\rm corr}^{(2)}$ with the basis set chosen in CCSD calculations and 
subtract from the converged $E_{\rm corr}^{(2)}$. The estimated $\Delta E$ in 
Table.\ref{engy_cct} is the sum of this difference and CCSD $\Delta E$.
This includes the correlation effects from $i$, $j$ and $k$ symmetries 
as well. For Ne, the estimated experimental value of correlation energy 
lies between the range 0.385 and 0.390 \cite{Sasaki-74,Bunge-70}.
Our coupled-cluster result, estimated value, is in excellent agreement.

\begin{table}[h]
\caption{Correlation energy from coupled-cluster. All the values are in 
         atomic units.}
\begin{ruledtabular}
\begin{tabular}{cccc}
 Atom & Active Orbitals &\multicolumn{2}{c}{$\Delta E$(ccsd)} \\
\hline                                                         
      &                 & Linear & Nonlinear                  \\
\hline
Ne & 17$s$10$p$10$d$9$f$9$g$8$h$    & -0.3783   & -0.3760     \\
   & 18$s$11$p$11$d$10$f$10$g$9$h$  & -0.3805   & -0.3782     \\
   & Estimated                      & -0.3905   & -0.3882     \\
                                                              \\
Ar & 17$s$11$p$11$d$9$f$9$g$9$h$    & -0.6884   & -0.6829     \\
   & 18$s$12$p$12$d$10$f$10$g$10$h$ & -0.7001   & -0.6945     \\
   & Estimated                      & -0.7258   & -0.7202     \\
                                                              \\
Kr & 22$s$13$p$11$d$9$f$9$g$9$h$    & -1.5700  & -1.5688      \\
   & 23$s$14$p$12$d$10$f$10$g$10$h$ & -1.6730  & -1.6716      \\
   & Estimated                      & -1.8480  & -1.8466      \\
                                                              \\
Xe & 23$s$14$p$12$d$10$f$10$g$10$h$ & -2.5500  & -2.5509      \\
   & 24$s$15$p$13$d$11$f$11$g$11$h$ & -2.6874  & -2.6881      \\
   & Estimated                      & -2.9973  & -2.9979      \\
\end{tabular}
\end{ruledtabular}
  \label{engy_cct}
\end{table}

The contributions to the correlation energy arising from the approximate  
triples are listed in Table.\ref{engy_t3}. As discussed in 
Section.\ref{corr_en}, the correlation energy diagrams corresponding to the
approximate triples are grouped into three classes. Out of these we have
selected a few: eight from 2p-2h and two each from 3p-1h and  1p-3h. In 
Table.\ref{engy_t3}, $\Delta E $  arising from these are listed. It is evident
from the table, the contribution from 1p-3h and 3p-1h are negative and adds 
to the magnitude of $\Delta E $. Whereas, the contribution from 2p-2h is 
positive and reduces the magnitude of $\Delta E $.
 
\begin{table}[h]
\caption{Correlation energy arising from the approximate triples 
         in the coupled-cluster theory. All the values are in atomic units.}
\begin{ruledtabular}
\begin{tabular}{ccccc}
 Atom & Basis size      &\multicolumn{3}{c}{$\Delta E$}               \\
      &                 & 2p-2h & 1p-3h & 3p-1h                       \\
\hline
Ne   & 18$s$11$p$11$d$10$f$10$g$9$h$  & 0.00672 & -0.00145 & -0.00164 \\
                                                                      \\
Ar   & 18$s$12$p$12$d$10$f$10$g$10$h$ & 0.00805 & -0.00066 & -0.00192 \\
                                                                      \\
Kr   & 22$s$13$p$11$d$9$f$9$g$9$h$    & 0.01546 & -0.00171 & -0.00305 \\
                                                                      \\
Xe   & 19$s$15$p$10$d$9$f$5$g$2$h$    & 0.02011 & -0.00148 & -0.00260 \\
\end{tabular}
\end{ruledtabular}
  \label{engy_t3}
\end{table}
%
%


\subsection{Dipole Polarizability}

  One constraint while using perturbed coupled-cluster theory to 
calculate dipole polarizability is the form of $\overline{D}$. It is a unitary
transformation of the dipole operator and expands to a non terminating series.
For the present calculations we consider the leading terms in 
${T^{(1)}}^\dagger \overline{D}$. That is, we use the approximation
\begin{eqnarray}
  {T^{(1)}}^\dagger \overline{D} &\approx &
  {T^{(1)}_1}^\dagger \left [ D +  D T^{(0)}_1 + D T^{(0)}_2 \right ] + 
              \nonumber \\
      &&  {T^{(1)}_2}^\dagger \left [ D T^{(0)}_2 + D T^{(0)}_1 \right ].
  \label{td_approx}
\end{eqnarray}
The ground state dipole polarizabilities of Ne, Ar, Kr and Xe calculated 
with this approximation  are given in Table.\ref{polarizibility}. 
Among the various terms, the first term ${T^{(1)}}^\dagger D$ subsumes 
contributions arising from  Dirac-Fock and random phase approximation. We can 
thus expect this term to have the most dominant contribution. This is 
evident in Table.\ref{polarizibility}, which shows that the contribution
from ${T^{(1)}}^\dagger D$ is far larger than the others.
The next leading term is ${T^{(1)}_1}^\dagger D T^{(0)}_2$. This is 
attributed to the larger values, compared to $T_1^{(0)}$,  
of $T_2^{(0)}$ cluster amplitudes. The dipole polarizability calculations with 
relativistic coupled-cluster theory involve two sets of cluster amplitudes. 
These are the $T^{(0)}$ and $T^{(1)}$ cluster amplitudes. As mentioned 
earlier, solving coupled-cluster equations is compute intensive. To minimize 
the computational costs, we optimize the basis parameters $\alpha_0$ and 
$\beta$ to contract the size of the orbital basis set. 

 One pattern discernible in the results is the better agreement between
the ${T^{(1)}}^\dagger D$ results and experimental data. The deviations
from the experimental data are large when we consider the total (CCSD) result.
For Ne, the deviation from experimental data is ~2\%, where as it is ~9\%
in the case of Xe atom. We attribute this to the approximation in 
Eq.(\ref{td_approx}) and partly to the basis set. To confirm this, however, 
requires detailed computations with higher order terms in the unitary 
transformation. This poses considerable computational challenges and shall 
be addressed in future publications.  We also implement the approximate 
triples excitation of the perturbed cluster amplitudes and contribution 
to $\alpha$ are listed in the table. 
\begin{table}[h]
\caption{\label{table6}Dipole polarizability of the ground state of
                           neutral rare-gas atoms (in a. u.).}
\begin{ruledtabular}
\begin{tabular}{ccccc}
 Contributions & Ne & Ar & Kr & Xe                                  \\
\hline
 ${T^{(1)}_1}^\dagger D$           &2.7108 &11.3330&17.2115&27.7427 \\
 ${T^{(1)}_1}^\dagger D T^{(0)}_1$ &0.0771 &0.0486 &0.0429 &-0.1495 \\
 ${T^{(1)}_1}^\dagger D T^{(0)}_2$ &-0.0703&-0.8264&-1.2721&-2.3286 \\
 ${T^{(1)}_2}^\dagger D T^{(0)}_1$ &-0.0004&-0.0001&0.0002 &0.0027  \\
 ${T^{(1)}_2}^\dagger D T^{(0)}_2$ &0.0053 &0.2490 &0.0439 &0.0786  \\
 Total(CCSD)                       &2.7225 &10.8041&16.0264&25.3459 \\
 Approx. triples                   &2.7281 &10.7360&16.0115&25.2974 \\
 Exp. values\footnotemark[1]       &2.670$\pm$0.005&11.070(7)&17.075&27.815
\end{tabular}
\end{ruledtabular}
\footnotetext[1]{\cite{hohm-90}.}
  \label{polarizibility}
\end{table}


\section{Conclusion}

  We have done a systematic study of the electron correlation energy of 
neutral inert gas atoms using relativistic MBPT and coupled-cluster theory. 
Our MBPT results are based on larger basis sets consisting of higher symmetries
than the previous works. Hence more reliable and accurate. Our study shows
that in heavier atoms Kr and Xe, the inner core electrons in $d$ shells 
dominates the electron correlation effects. This ought to be considered in 
high precision properties calculations. For example, the EDM calculations of
Xe arising from nuclear Schiff moment. The dipole polarizability 
calculated with the perturbed coupled-cluster show systematic deviation
from the experimental data. However, the leading term is in good agreement. 
The deviations might decrease when higher order terms are incorporated in 
the calculations. From these results, it is evident that the basis set 
chosen is of good quality and appropriate for precision calculations.


\begin{acknowledgments}
We wish to thank S. Chattopadhyay, S. Gautam, K. V. P. Latha, B. Sahoo and 
S. A. Silotri  for useful discussions. The results presented in the paper
are based on computations using the HPC cluster of the Center for 
computational Material Science, JNCASR, Bangalore. 
\end{acknowledgments}


\end{document}